\documentclass{PoS}

\pdfoutput=1 

\usepackage{graphicx}
\usepackage{graphics}
\usepackage{amsmath}
\usepackage{bm}% bold math

\newcommand{\hermes}{\textsc{Hermes }}
\newcommand{\compass}{\textsc{Compass }}
\newcommand{\belle}{\textsc{Belle }}
\newcommand{\babar}{\textsc{BaBar }}

\newcommand{\gmctrans}{\textsc{gmctrans }}

\newcommand{\p}{\perp}

\newcommand{\smarrow}{\mbox{\raisebox{-4.5pt}[0pt][0pt]{$\hspace{-1pt}
      \vec{\phantom{v}}$}}}

\title{Phenomenology of unpolarized TMDs from Semi-Inclusive DIS data}

\ShortTitle{Phenomenology of unpolarized TMDs from SIDIS data}


\author{\speaker{Andrea Signori}$^{\ a,b}$, Alessandro Bacchetta,$^{c,d}$ and Marco Radici$^{d}$ \\
\\
\llap{$^a$} Department of Physics and Astronomy, VU University Amsterdam \\ De Boelelaan 1081, NL-1081 HV Amsterdam, the Netherlands \\
\llap{$^b$} Nikhef \\ Science Park 105, NL-1098 XG Amsterdam, the Netherlands \\
\llap{$^c$} INFN Sezione di Pavia \\ via Bassi 6, I-27100 Pavia, Italy \\
\llap{$^d$} Dipartimento di Fisica, Universit\`a di Pavia \\ via Bassi 6, I-27100 Pavia, Italy \\
\\
E-mail: \email{asignori@nikhef.nl}, \email{alessandro.bacchetta@unipv.it}, \email{marco.radici@pv.infn.it}}

\abstract{
We discuss the dependence on the flavor of the intrinsic transverse momentum in unpolarized transverse-momentum-dependent distribution functions (TMD PDFs) and fragmentation functions (TMD FFs) analyzing data of semi-inclusive deep-inelastic scattering (SIDIS) released by the \hermes collaboration.  
Adopting a flavor-dependent Gaussian model in the transverse momentum and neglecting QCD evolution, we find interesting evidences concerning the flavor dependence in TMD FFs, whereas the indications are weaker in TMD PDFs and deserve further investigations. 
Inclusions of new data sets of SIDIS, e$^+$e$^-$ annihilations and and Drell-Yan (DY) events require a proper treatment of QCD evolution. We try to get constraints on the non-perturbative Sudakov factor from e$^+$e$^-$ annihilations into hadrons.
}

\FullConference{XXII. International Workshop on Deep-Inelastic Scattering and Related Subjects \\
		 28 April - 2 May 2014 \\
		 Warsaw, Poland}
		 
\begin{document}

%%%%%%%%%%%%%%%%%%%%%%%%%%%%%%%%%%%%%%%%%%%%%%%%%%%%%%%%%%%%%%%%%%%%%%%%%%%%%%%%%%%%%%%%
\section{Introduction and motivation}
%TMD PDFs and TMD FFs (collectively referred as TMDs) are functions of the longitudinal and transverse momentum of partons, mathematical tools helpful to map the structure of hadrons in three-dimensional momentum space. 
TMDs are not phenomenologically well known. For example, even in the simple {\em unpolarized} distribution $f_1(x,k_\p^2)$ (i.e., the distribution of partons with flavor $a$ summed over their polarization and averaged over the polarization of the parent hadron) the transverse momentum dependence is poorly known. This work refers to one of the first phenomenological investigations (see Refs.~\cite{Signori:2013mda},~\cite{Signori:2013gra}) of the unpolarized TMDs, addressing their flavor structure in the transverse momentum.
%The SIDIS data published by the \hermes collaboration are ideal to address this issue, since they refer to SIDIS off protons and deuterons with charge-separated pions and kaons in the finale state, with multidimensional binning in the kinematic variables $\xbj$, z and $Q^2$. SIDIS data released by the \compass collaboration will be considered in the future.\\
%included in forthcoming analyses together with a proper treatment of QCD evolution to separate intrinsic and perturbative components of partonic transverse momentum. \\
For a list of physical motivations underlying this study see Ref.~\cite{Signori:2013mda}. Above all, it is important to notice that
%We believe there are stringent motivations to study the flavor dependence of intrinsic transverse momentum of partons.
% collinear distributions
%We know with good accuracy and precision that the unpolarized collinear PDFs $f_1^a(x)$ and FFs $D_1^a(z)$ strongly depend on the parton flavor $a$ (cita una pdf e DSS) (see, e.g., Refs.~\cite{Forte:2013wc,Gao:2013xoa,Owens:2012bv,Ball:2012cx,Martin:2009iq,JimenezDelgado:2008hf}). For this reason we believe it's natural to check if there is a dependence on the flavor also in the transverse-momentum-dependent part. 
%Other gorups alredy investigated this topic
% lattice calculations for TMD PDFs
%through lattice QCD (see Ref.~\cite{Musch:2010ka}) and model calculations:
% model calculations for TMD PDFs and TMD FFs
%for example, the chiral quark soliton model, the diquark spectator model, the statical approach to distribution functions, the NJL-jet model for fragmentation functions (pair the references see~\cite{Bacchetta:2008af,Bacchetta:2010si,Wakamatsu:2009fn,Efremov:2010mt,Bourrely:2010ng,Matevosyan:2011vj,Schweitzer:2012hh}) predict different transverse-momentum behaviors for different flavors. 
%Other calculations, instead, do not (see \cite{Pasquini:2008ax,Lorce:2011dv,Avakian:2010br}). 
% previous (SIDIS) fits
%There is also a fit of SIDIS data (see ref to JLab Hall C Collaboration) addressing preliminary the topic.
% data sets are different!
%Last but not least, we know that 
the SIDIS data sets for different combinations of target and detected hadrons are different. Introducing the flavor dependence in the TMD part of distribution and fragmentation functions allows us to {\em theoretically} discriminate the available configurations. This is not possible in the simple flavor-independent Gaussian ansatz.

%%%%%%%%%%%%%%%%%%%%%%%%%%%%%%%%%%%%%%%%%%%%%%%%%%%%%%%%%%%%%%%%%%%%%%%%%%%%%%%%%%%%%%%%
\section{Unpolarized SIDIS and flavor-dependent Gaussian TMDs}
In one-hadron inclusive DIS a lepton scatters off a nucleon $N$ and one hadron $h$ is identified in the final state:
\begin{equation}
\ell + N \to \ell' + h + X \ ,
\label{sidis}
\end{equation}
where $\ell$, $\ell'$ denote the incoming and outgoing lepton respectively.
%% definition of transverse momenta
The light-cone transverse momenta involved in the process are:\\

\noindent
\begin{tabular}{ll}
Momentum  & Physical meaning \\
\hline
$\bm{k}_\p$ & intrinsic partonic transverse momentum  
\\ 
$\bm{P}_\p$ & transverse momentum of final hadron w.r.t. fragmenting parton
\\
$\bm{P}_{hT}$ & transverse momentum of final hadron w.r.t. virtual photon
\end{tabular} \\

We consider unpolarized scattering integrated over the azimuthal angle $\phi_h$ of the detected hadron in the limit of small transverse momenta and at leading twist (respectively, $\bm{P}_{hT}^2 \ll Q^2$ and $M^2 \ll Q^2$, being $Q^2$ the hard scale of the process and $M$ the mass of the target hadron). 
As a first investigation, we adopt the parton model picture performing a leading order QED analysis (one-photon exchange approximation) neglecting any modification that can be induced by QCD evolution. This is possible because of the limited range in $Q^2$ (see Refs.~\cite{Airapetian:2012ki},~\cite{Signori:2013mda}).

The \hermes collaboration released SIDIS multiplicities (see Ref.~\cite{Airapetian:2012ki}), the differential number of hadrons produced per DIS event. Multiplicity is defined as the ratio of the SIDIS and the DIS cross sections and, at leading twist, is proportional to the unpolarized transverse structure function $F_{UU,T}$ (see Ref.~\cite{Bacchetta:2006tn}):
\begin{equation}
m_N^h (x,z,Q^2,\bm{P}_{hT}^2) = \frac{d^{(4)} \sigma_N^h / dx dQ^2 dz d\bm{P}_{hT}^2 }{ d^{(2)}\sigma_{\text{DIS}} / dx dQ^2 }\ \sim \ F_{UU,T}(x,z,Q^2,\bm{P}_{hT}^2) .
\label{e:multiplicity}
\end{equation}
%
%\begin{equation}
%\frac{d^{(4)}\sigma}{dx\ dQ^2 \ dz \ d\bm{P}_{hT}^2} = \frac{\pi \alpha^2}{x Q^4} \biggl[ 1+ \biggl( 1-\frac{Q^2}{x(s-M^2)} \biggr) \biggr] \ F_{UU,T}(x,z,Q^2,\bm{P}_{hT}^2)\ ,
%\label{e:cross_LT_LO}
%\end{equation} 
%
$F_{UU,T}$ is expressed in terms of a convolution of a TMD PDF and a TMD FF relying on the factorized formula for low transverse momentum SIDIS:
\begin{equation} 
\label{F_UUT_simpl}
%\hspace{-3mm}
F_{UU ,T} (x,z,Q^2,\bm{P}_{hT}^2) =  \sum_a e_a^2\ \bigg[f_1^{a,N}(x,\bm{k}_{\perp}^2,Q^2) \otimes D_1^{a \to h}(z,\bm{P}_{\perp}^2, Q^2) \bigg] \ ,
\end{equation}
where $a$ is the flavor index\footnote{In summing over the flavor index up and down contributions will consist of two parts: one for valence quarks and one for sea quarks.} 
and the convolution is defined in Ref.~\cite{Bacchetta:2006tn}. A more extensive description of the assumptions, the notation, and the general theoretical framework is available in Ref.~\cite{Signori:2013mda}.

%%%%%%%%%%%%%%%%%%%%%%%%%%%%%%%%%%%%%%%%%%%%%%%%%%%%%%%%%%%%%%%%%%%%%%%%%%%%%%%%%%%%%%%%
%\section{Flavor-dependent unpolarized Gaussian TMDs}
The flavor-dependent Gaussian hypothesis consists in assuming flavor-dependent Gaussian behavior in the transverse momentum in the unpolarized distribution function $f_1^{a,N}$ and fragmentation function $D_1^{a \to h}$:

\begin{align}
f_1^{a,N}(x,Q^2,\bm{k}_\perp^2) = \frac{f_1^{a,N}(x,Q^2)}{\pi \langle \bm{k}_{\perp,a}^2 \rangle}e^{-\bm{k}_\perp^2/\langle \bm{k}_{\perp,a}^2 \rangle} , \ \ \ \ \ 
D_1^{a \to h}(z,Q^2,\bm{P}_\perp^2) = \frac{D_1^{a \to h}(z,Q^2)}{\pi \langle \bm{P}_{\perp,a \to h}^2 \rangle}e^{- \bm{P}_\perp^2/\langle \bm{P}_{\perp,a \to h}^2 \rangle}\ .
\label{e:fldep_gauss}
\end{align} 
%
%Introducing this assumption the unpolarized structure function reads:
Introducing this assumption the multiplicity reads:
%\begin{equation} 
%\hspace{-3mm}
%F_{UU ,T} =  \sum_a e_a^2\ f_1^a(x,Q^2) D_1^{a \to h}(z,Q^2)\ \bigg[ \frac{e^{-\bm{k}_\perp^2/\langle \bm{k}_{\perp,a}^2 \rangle}}{\pi \langle \bm{k}_{\perp,a}^2 \rangle} \otimes \frac{e^{- \bm{P}_\perp^2/\langle \bm{P}_{\perp,a \rightarrow h}^2 \rangle}}{\pi \langle \bm{P}_{\perp,a \rightarrow h}^2 \rangle} \bigg] \ .
%\label{e:F_UUT_fldep}
%\end{equation}
%
%Each convolution in Eq. \ref{e:F_UUT_fldep} results in a Gaussian function in $\bm{P}_{{hT}}$
%\begin{equation}
%\bigg[ \frac{1}{\pi \langle \bm{k}_{\perp,a}^2 \rangle}e^{-\bm{k}_\perp^2/\langle \bm{k}_{\perp,a}^2 \rangle} \otimes \frac{1}{\pi \langle \bm{P}_{\perp,a \rightarrow h}^2 \rangle}e^{- \bm{P}_\perp^2/\langle \bm{P}_{\perp,a \rightarrow h}^2 \rangle} \bigg] = \frac{x}{\pi \langle \bm{P}_{hT,a}^2 \rangle}e^{-\bm{P}_{hT}^2/\langle \bm{P}_{hT,a}^2 \rangle} \ ,
%\label{e:Gauss_con}
%\end{equation}
%
%where the relation between the three variances is:
%
%\begin{equation}
%\langle \bm{P}_{hT,a}^2 \rangle = z^2 \langle \bm{k}_{\perp,a}^2 \rangle + \langle \bm{P}_{\perp,a \rightarrow h}^2 \rangle \ .
%\label{e:fldep_transvmom_rel}
%\end{equation}
%Through Eq. \ref{e:fldep_transvmom_rel}, peculiar of the Gaussian formulation, the measured transverse momentum $\langle \bm{P}_{hT,a}^2 \rangle$ becomes the experimental handle to access the average square values of the intrinsic transverse momenta $\bm{k}_\perp$ and $\bm{P}_\perp$.
%
%Evaluating the cross sections using the flavor-dependent Gaussian hypothesis we obtain the following expression:
\begin{equation}
\begin{split}
m_{N}^h (x,z,Q^2,\bm{P}_{hT}^2) &= \frac{ \pi }{ \sum_{a} e_a^2 f_{1}^{a,N} (x,Q^2) }  \\
& \times \sum_{a} \biggl[ e_a^2 f_{1}^{a,N} (x,Q^2) D_{1}^{a \to h} (z,Q^2)\ \frac{ e^{ - \frac{ \bm{P}_{hT}^2 }{ z^2 \langle \bm{k}_{\perp,a}^2 \rangle + \langle \bm{P}_{\perp,a \to h}^2 \rangle }} }{ \pi (z^2 \langle \bm{k}_{\perp,a}^2 \rangle + \langle \bm{P}_{\perp,a \to h}^2 \rangle) } \biggr]  \ ,
\label{e:FDmult}
\end{split}  
\end{equation} 
where the Gaussian functions in $\bm{P}_{{hT}}$ result from the convolution of $f_1^{a,N}$ and $D_1^{a \to h}$.
The multiplicity is not a Gaussian in $\bm{P}_{{hT}}$, because it is a summation of Gaussians functions. This is important, since it experimentally deviates from the Gaussian functional form.

%%%%%%%%%%%%%%%%%%%%%%%%%%%%%%%%%%%%%%%%%%%%%%%%%%%%%%%%%%%%%%%%%%%%%%%%%%%%%%%%%%%%%%%%
\section{Assumptions concerning average transverse momenta}
We adopt a simple framework for flavor dependence, introducing different widths for up-valence, down-valence and sea quarks, with a dependence on the longitudinal momentum fraction $x$. The latter is driven by data analysis, but above all it is suggested by theoretical considerations and model calculations (see Ref.~\cite{Signori:2013mda}):
%(see,  e.g.,~\cite{Bacchetta:2008af,Bacchetta:2010si,Wakamatsu:2009fn,Efremov:2010mt,%Bourrely:2010ng,Matevosyan:2011vj,Schweitzer:2012hh,%Pasquini:2008ax,Lorce:2011dv,Avakian:2010br}),
%and similarly do parametrizations of light-front wave functions (see,
%e.g.,~\cite{Brodsky:2000ii,Hwang:2007tb,Gutsche:2013zia}).  
%
%We choose the following functional form for the average square transverse momentum of flavor $a$:
\begin{align} 
\big\langle \bm{k}_{\p,a}^2 \big\rangle (x) = 
\big\langle \hat{\bm{k}}_{\p,a}^2 \big\rangle \;  
\frac{(1-x)^{\alpha} x^{\sigma} }{ (1-\hat{x})^{\alpha} \hat{x}^{\sigma} } \, ,
\label{e:kT2_kin}
&&
\text{where }
\big\langle \hat{\bm{k}}_{\p,a}^2 \big\rangle\equiv \big \langle \bm{k}_{\p,a}^2
\big \rangle
(\hat{x}),
\text{ and }
\hat{x}=0.1.
\end{align} 
$\langle \hat{\bm{k}}_{\p,a}^2 \rangle$, $\alpha$, $\sigma$ are free parameters:
$\alpha$ and $\sigma$ are flavor-independent, whereas we have three flavor-dependent normalizations, $\langle \hat{\bm{k}}_{\p,a}^2 \rangle$ for $a = u_v,\, d_v,\, {\rm sea}$. 
%In total, we use five different parameters to describe all TMD PDFs.
%Since the present data have a limited coverage in $x$, we found no need of more sophisticated choices.  
Concerning TMD FFs, we distinguish three favored process and one class of unfavored processes, assuming charge conjugation and isospin symmetry:
%As for  TMD FFs, fragmentation processes in which the fragmenting parton is in the valence content of the detected hadron are usually defined {\em favored}. Otherwise the process is classified as {\em unfavored}. The biggest difference between the two classes is the number of $q\bar{q}$ pairs excited from the vacuum in order to produce the detected hadron: favored processes involve the creation of at most one $q\bar{q}$ pair. If the final hadron is a kaon, we further distinguish a favored process initiated by a strange quark/antiquark from a favored process initiated by an up quark/antiquark. 
%For simplicity, we assume charge conjugation and isospin symmetries. The latter is often imposed also in the parametrization of collinear FFs~\cite{Hirai:2007cx}, but not always~\cite{deFlorian:2007aj}. In practice, we consider four different Gaussian shapes: 
\begin{gather}
\big\langle \bm{P}^2_{\p,u \smarrow \pi^+} \big\rangle = \big \langle
\bm{P}^2_{\p,\bar{d} \smarrow \pi^+} \big \rangle = \big \langle \bm{P}^2_{\p,\bar{u} \smarrow \pi^-} \big \rangle =  \big \langle \bm{P}^2_{\p,d \smarrow \pi^-}\big \rangle \equiv \big \langle \bm{P}^2_{\p, {\rm fav}} \big \rangle \, ,  
\label{e:favored}  
\\
\big \langle \bm{P}^2_{\p,u \smarrow K^+} \big \rangle =  \big \langle \bm{P}^2_{\p,\bar{u} \smarrow K^-} \big \rangle \equiv \big \langle \bm{P}^2_{\p, {uK}} \big \rangle \, ,
\label{e:uK}  
\\
\big \langle \bm{P}^2_{\p,\bar{s} \smarrow K^+} \big \rangle = \big \langle \bm{P}^2_{\p,s \smarrow K^-}\big \rangle \equiv \big \langle \bm{P}^2_{\p, {sK}} \big \rangle \,  ,
\label{e:sK}  
\\
\big \langle \bm{P}^2_{\p,\text{all others}} \big \rangle 
%\big \langle P_{\p,\bar{u} \smarrow \pi^+} \big \rangle =
%\big \langle P_{\p,\bar{d} \smarrow \pi^-} \big \rangle = \big \langle P_{\p, u \smarrow \pi^-} \big \rangle = 
%\big \langle P_{\p, (s,b,c) \smarrow \pi^+} \big \rangle = \big \langle P_{\p, (s,c,b) \smarrow \pi^-} 
\equiv  \big \langle \bm{P}^2_{\p, {\rm unf}} \big \rangle \, .
\label{e:unfavored} 
\end{gather} 
%The last assumption is to keep the number of parameters under control, though it is clear that unfavored fragmentation into kaons is different from unfavored fragmentation into pions. 
As for the previous case, we introduce a dependence on the longitudinal momentum fraction $z$ in the variance of the Gaussian TMD FFs:
\begin{align}  
\big \langle \bm{P}_{\p,a \smarrow h}^2 \big \rangle (z) &= \big \langle
\hat{\bm{P}}_{\p,a \smarrow h}^2 \big \rangle 
\frac{ (z^{\beta} + \delta)\ (1-z)^{\gamma} }{ (\hat{z}^{\beta} + \delta)\
  (1-\hat{z})^{\gamma} } \, 
&&
\text{where }
\big\langle \hat{\bm{P}}_{\p,a \smarrow h}^2 \big\rangle\equiv \big \langle
\bm{P}_{\p,a \smarrow h}^2
\big \rangle
(\hat{z}),
\text{ and }
\hat{z}=0.5.
\label{e:PT2_kin}
\end{align}  
The kineamtic parameters $\beta$, $\gamma$ and $\delta$ are flavor-independent, whereas we have one normalization for each class of fragmentation processes.

%%%%%%%%%%%%%%%%%%%%%%%%%%%%%%%%%%%%%%%%%%%%%%%%%%%%%%%%%%%%%%%%%%%%%%%%%%%%%%%%%%%%%%%%
\section{Phenomenology}

%%%%%%%%%%%%%%%%%%%%%%%%%%%%%%%%%%%%%%%%%%%%
%\subsection{Data selection and fit procedure}

%The full kinematic coverage provided by Hermes is $(x,Q^2)$ ranging from about $(0.04,1.25\ \text{ GeV}^2)$ to about $(0.4,9.2\
%\text{ GeV}^2)$, $z$ from 0.1 to 0.9, $P_{h \perp}$ from 0.1 to 1. 
%The analysis has been restricted to particular kinematic regions, looking for a good agreement between the data and the model.
%Cuts imposed on $z, Q^2$ are driven by the matching between values of collinear multiplicities and Eq. \ref{e:FDmult} integrated over transverse momentum. Cuts on $\bm{P}_{h\perp}$ values are related to the applicability of the TMD formalism.

%Summarizing:
We fit the \hermes data sets related to proton and deuteron targets and pions and kaons in the final state, with multidimensional binning in $x$, $z$ and $Q^2$. \compass data will be considered in the future.
Since the $Q^2$ range is narrow, we can carry out the analysis at a fixed $Q^2 = 2.4\ \text{GeV}^2$ in Eq.~\eqref{e:FDmult}, neglecting any effect from QCD evolution.
We fit ${\cal M}=$ 200 Monte Carlo replicas of the original data set, in order to get ${\cal M}$ best values for each parameter. 
%This allows, for example, a deeper insight into the physical obsvervables computable from the fit results.
A detailed description of the kinematic cuts on the data and the Monte Carlo fit procedure is given in Ref.~\cite{Signori:2013mda}.

We perform four fits with different assumptions. 
The first one, conventionally named {\em default} fit, is based on the less restrictive assumption for kinematic cuts. 
In the second one, we further exclude data with $Q^2 < 1.6$ GeV$^2$.  
The third scenario corresponds to neglecting kaons in the final state. 
The last scenario is a fit of the default selection using a flavor-independent Gaussian ansatz. 
Here we will shortly present results related to the first and the fourth fits, the others being available in Ref.~\cite{Signori:2013mda}.

In the default fit the average global $\chi^2/{\rm  d.o.f.}$ is $1.63 \pm 0.13$. 
The quality of the fit is not excellent, but we need to take into account that the description of the multiplicities is difficult already in the collinear case (see Ref.~\cite{Signori:2013mda}). 
In the flavor-independent case the agreement is slightly worse, with an average global $\chi^2/{\rm  d.o.f.}$ of $1.72 \pm 0.11$. This is because with the flavor-independent model we fit different data sets with the same model function.

\begin{figure}
\centering
\begin{tabular}{ccc}
\includegraphics[width=7.0cm]{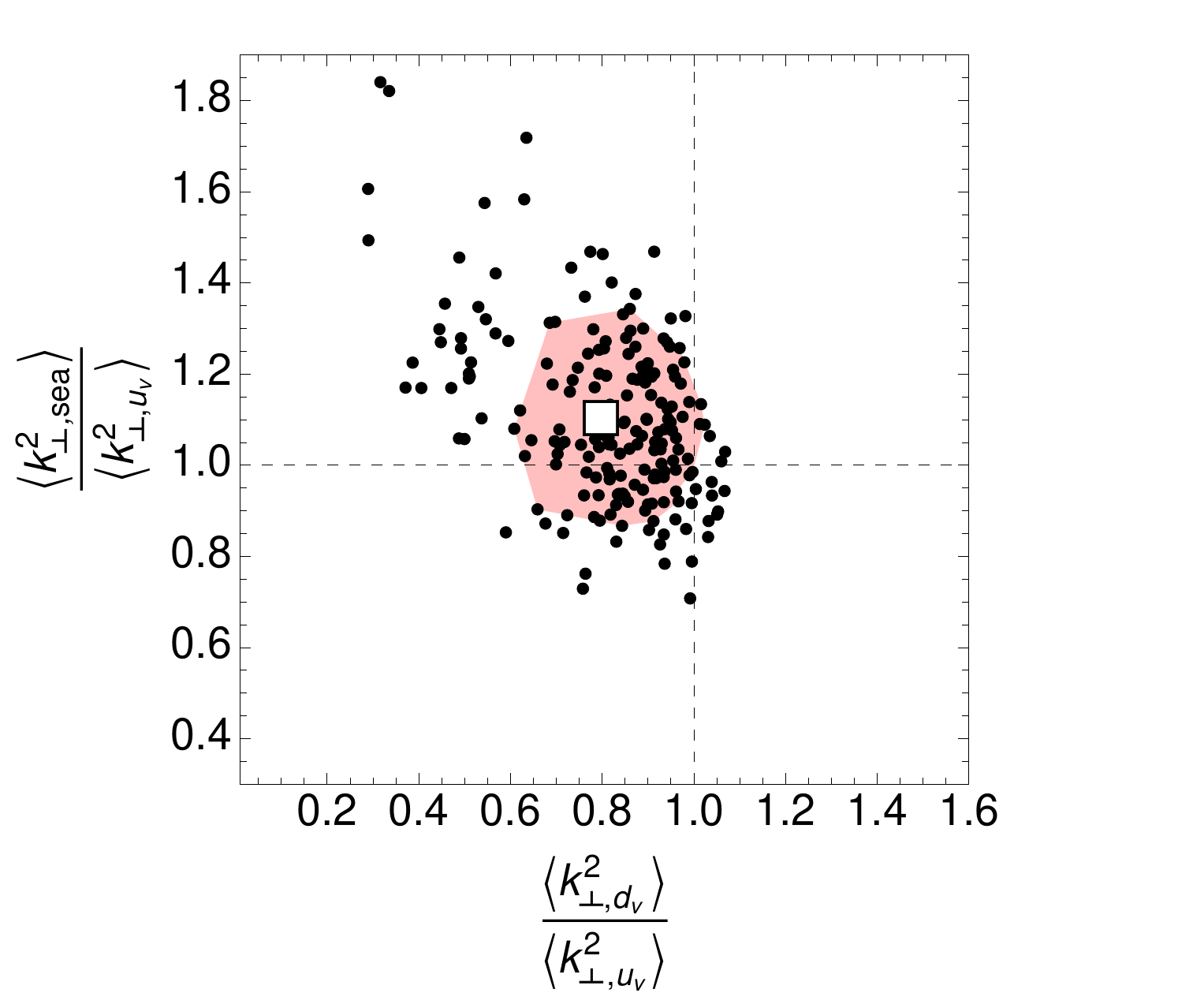}
&\hspace{0.001cm}
&
\includegraphics[width=7.0cm]{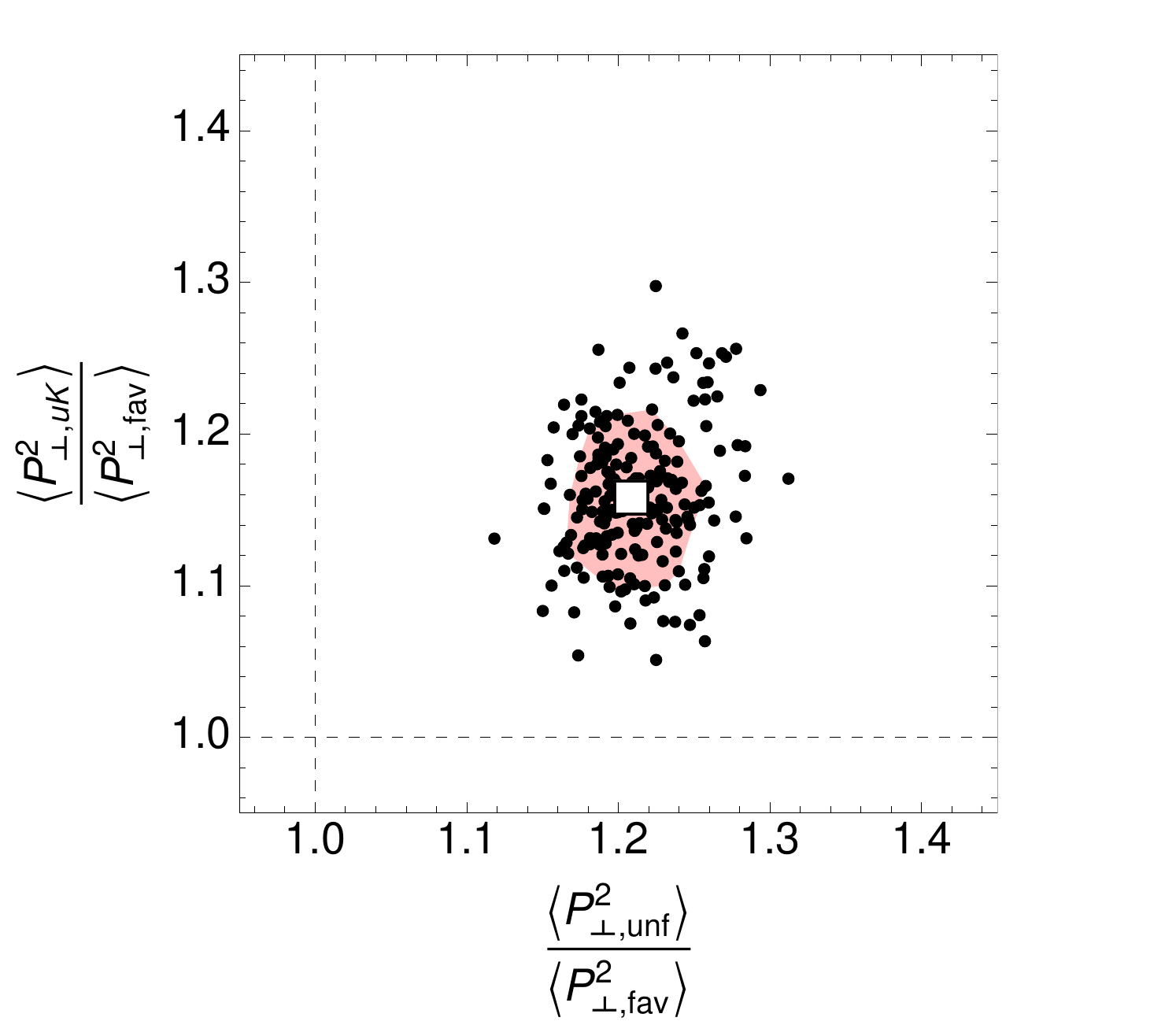}
\\
(a) && (b)
\end{tabular}
\caption{(a) Distribution of the values of the ratios 
$\langle \bm{k}^2_{\p, d_v} \rangle / \langle \bm{k}^2_{\p, u_v}\rangle$ vs. 
$\langle \bm{k}^2_{\p, {\rm sea}}\rangle / \langle \bm{k}^2_{\p, u_v}\rangle$ obtained
from fitting 200 replicas of the original data points. The white squared box indicates the center of the 68\% confidence
interval for each ratio.
The shaded area represents the two-dimensional 68\% confidence
region around the white box. 
The dashed lines correspond to the ratios being unity; their crossing
point corresponds to the result with no flavor dependence. For most of the
points,  
$\langle \bm{k}^2_{\p, d_v}\rangle < \langle \bm{k}^2_{\p, u_v}\rangle < \langle
\bm{k}^2_{\p, {\rm sea}}\rangle$.  
(b) Same as previous panel, but for the distribution of the values of the
ratios  $\langle \bm{P}^2_{\p, {\rm
    unf}}\rangle / \langle \bm{P}^2_{\p, {\rm fav}}\rangle$  vs. 
$\langle \bm{P}^2_{\p, u  K}\rangle / \langle \bm{P}^2_{\p, {\rm
    fav}}\rangle$. 
For all points, 
$\langle \bm{P}^2_{\p, {\rm fav}}\rangle < \langle \bm{P}^2_{\p, {\rm
    unf}}\rangle \sim \langle \bm{P}^2_{\p, u K}\rangle$. 
}
\label{f:DoverU_SoverU_default}
\end{figure}

In Tab.~\ref{t:fd_PDFs_par} and Tab.~\ref{t:fd_FFs_norm} we list the average best values for some~\footnote{for a complete list please see Ref.~\cite{Signori:2013mda}.} of the fit parameters related to TMD PDFs and TMD FFs respectively.
%The most interesting plots to investigate the differences between the flavor-dependent and the flavor-independent Gaussian approximations are shown in Fig.~\ref{f:DoverU_SoverU_default}.
In the left panel of Fig.~\ref{f:DoverU_SoverU_default}, we compare the ratio
$\langle \bm{k}_{\p, d_v}^2 \rangle / \langle \bm{k}_{\p, u_v}^2 \rangle$
vs. $\langle \bm{k}_{\p, {\rm sea}}^2 \rangle / \langle \bm{k}_{\p, u_v}^2
\rangle$ calculated within the 200 replicas. 
The dashed lines correspond to unitary ratios and divide the plane into four sectors. 
For many replicas, the values lie in the upper left quadrant, meaning that 
$\langle \bm{k}_{\p, d_v}^2 \rangle < \langle \bm{k}_{\p, u_v}^2 \rangle < \langle \bm{k}_{\p, {\rm sea}}^2 \rangle$.
On average, $d_v$ is about 20\% narrower than  $u_v$, which is about 10\% narrower than the sea. 
The crossing of the dashed lines corresponds to the flavor-independent configuration, in which all transverse momenta are equal. 
This point lies at the limit of the 68\% confidence region, meaning that the flavor-independent configuration is not statistically ruled out by the Monte Carlo fit.
In the right panel of Fig.~\ref{f:DoverU_SoverU_default}, we compare the ratio
$\langle \bm{P}_{\p, {\rm unf}}^2 \rangle / \langle \bm{P}_{\p, {\rm fav}}^2 \rangle$ vs. $\langle \bm{P}_{\p, u K}^2 \rangle / \langle \bm{P}_{\p, {\rm fav}}^2 \rangle$ .
On average, the width of unfavored and $u \to K$ fragmentations are about 20\% larger than the widht of favored ones.
All points are concentrated in the upper right quadrant: we have the {\em clear} outcome that $\langle \bm{P}_{\p, {\rm fav}}^2 \rangle < \langle \bm{P}_{\p, {\rm unf}}^2 \rangle \sim \langle \bm{P}_{\p, u K}^2 \rangle$ and that the flavor-independent configuration is well outside the 68\% confidence region.
Results from a flavor-independent Gaussian fit of \hermes and \compass data are also available in Ref.~\cite{Anselmino:2013lza}.

%%%%%%%%%%%  ratios of transverse momenta  %%%%%%%%%%%%%
%%%%%%%%%%%%%%%%%%%%%%%%%%%%%%%%%%%%%

%%%%%%%%%%%%   Tab PDFs parameters  %%%%%%%%%%%%%%
% transverse momenta - normalization
\begin{table}
\small
  \centering
  \begin{tabular}{|c|c|c|c|}
  \hline
  \multicolumn{4}{|c|}{Parameters for TMD PDFs} \\
  \hline
  \hline 
	&  $\big \langle \hat{\bm{k}}_{\p, d_v}^2 \big \rangle$ [GeV$^2$]  
	&  $\big \langle \hat{\bm{k}}_{\p, u_v}^2 \big \rangle$ [GeV$^2$]  
	&  $\big \langle \hat{\bm{k}}_{\p, {\rm sea}}^2 \big \rangle$ [GeV$^2$] 
%	&  $\alpha$ %(random)  
%	&  $\sigma$ %(random)  

\\
  \hline
  \hline
  Default   &   $0.30\pm 0.17$   &   $0.36\pm 0.14$   &   $0.41\pm 0.16$  \\ \hline % &   $0.95\pm 0.72$   &   $-0.10\pm 0.13$    
  Flavor-indep.	  &   $0.30\pm 0.10$   &   $0.30\pm 0.10$   &   $0.30\pm 0.10$  \\ % &   $1.03\pm 0.64$   &   $-0.12\pm 0.12$  \\
\hline
\end{tabular}
\caption{68\% confidence intervals of average mean square transverse momenta for TMD PDFs in the different scenarios.}
\label{t:fd_PDFs_par}
\end{table}

\begin{table}[h]
\small
  \centering
  \begin{tabular}{|c|c|c|c|c|}
  \hline
  \multicolumn{5}{|c|}{Parameters for TMD FFs} \\
  \hline
  \hline 
     &  $\big \langle \hat{\bm{P}}_{\p, \text{fav}}^2 \big \rangle $ [GeV$^2$] 
     &  $\big \langle \hat{\bm{P}}_{\p, \text{unf}}^2 \big \rangle $ [GeV$^2$]
     &  $\big \langle \hat{\bm{P}}_{\p, s K}^2 \big \rangle $  [GeV$^2$]
     &  $\big \langle \hat{\bm{P}}_{\p, u K}^2 \big \rangle $  [GeV$^2$] \\
\hline
\hline
  Default  &  $0.15\pm 0.04$  &  $0.19\pm 0.04$  &  $0.19\pm 0.04$  &  $0.18\pm0.05$  \\
\hline
  Flavor-indep.  &  $0.18\pm 0.03$  &  $0.18\pm 0.03$  &  $0.18\pm 0.03$  &  $0.18\pm 0.03$  \\
\hline
\end{tabular}
\caption{68\% confidence intervals of average mean square transverse momenta for TMD FFs in the different scenarios.}
\label{t:fd_FFs_norm}
\end{table}
\section{Extensions}
In all the configurations we observe a strong anticorrelation between transverse momenta in distribution and fragmentation functions. 
This is not surprising, since the width of the observed $\bm{P}_{hT}$ distribution is given in Eq.~\eqref{e:FDmult} as a summation of partonic transverse momenta. To better constrain the transverse momenta in TMD PDFs and FFs it will be essential to include also data from electron-positron annihilations and Drell--Yan processes. In facing this task it is fundamental to introduce QCD evolution, to distinguish perturbative from non-perturbative components in partonic transverse momenta (see, e.g., Ref.~\cite{Collins:2011zzd} and~\cite{Echevarria:2012pw}).
Assuming to describe the soft initial-scale transverse momenta with flavor-dependent Gaussian distributions, we still do not know the compatible values for the parameters modelling the soft contributions to the evolution. 
%the values of $g_2$ compatible with our flavor-dependent Gaussian ${\cal{F}}_{PDF}^{NP}$ and ${\cal{F}}_{FF}^{NP}$.
For this reason, we calculate the transverse-momentum-dependent cross section for e$^+$e$^-$ annihilations into one and two hadrons, aiming to constrain the most physical configurations in the evolution 
%$g_2$ and $b_\text{max}$ 
from the comparison with forthcoming data (to be released by the \belle and \babar collaborations). At the same time, we want to pin down the most physical flavor-dependent replicas of Gaussian TMD FFs out of the $200$ equivalent ones fitted on the \hermes data.
Once we fix the parameters for the evolution
%values for $g_2$ and $b_\text{max}$ 
compatible with our extracted TMD FFs, we will have a consistent framework to perform a global fit of different data sets.

%%%%%%%%%%%%%%%%%%%%%%%%%%%%%%%%%%%%%%%%%%%%%%%%%%%%%%%%%%%%%%%%%%%%%%%%%%%%%%%%%%%%%%%%
\section{Conclusions}

% impact on HEP
We moved the first steps in the process of investigating the flavor dependence of partonic transverse momentum through unpolarized TMDs. 
The impact of this study, however, will go beyond unpolarized distributions: it will also affect the extractions of the polarized ones, because the unpolarized TMDs enter the denominator of spin asymmetries. Moreover, it will be interesting to test its effect on Monte Carlo generators sensitive to the intrinsic transverse momentum of partons (like \gmctrans, Herwig, Pythia, ResBos, Cascade). 
To get a better phenomenological perspective we need to include more data in a framework which properly takes into account QCD evolution. Further papers will describe the achievements made along this way. 
This work is part of the program of the Stichting voor Fundamenteel Onderzoek der Materie (FOM), supported by the Nederlandse Organisatie voor Wetenschappelijk Onderzoek (NWO).

%%%%%%%%%%%%%%%%%%%%%%%%%%%%%%%%%

%%%%%%%%%%%%%%%%%%%%%%%%%%%%%%%%%

\end{document}